\documentclass[conference,a4paper,10pt,times]{IEEEtran}
\IEEEoverridecommandlockouts
\usepackage{cite}
\usepackage{amsmath,amssymb,amsfonts}
\usepackage{algorithmic}
\usepackage{graphicx}
\usepackage{textcomp}
\usepackage{xcolor}
\def\BibTeX{{\rm B\kern-.05em{\sc i\kern-.025em b}\kern-.08em
    T\kern-.1667em\lower.7ex\hbox{E}\kern-.125emX}}
\usepackage{tabularx}
\usepackage{fancyhdr}
\fancypagestyle{firstpage}{
  \fancyhf{}
  
  \fancyhead[C]{Paper accepted at the 4th IEEE European Symposium on Security and Privacy (EuroS\&P'19)}     
}
\begin{document}

\title{Understanding eWhoring}

\author{\IEEEauthorblockN{Alice Hutchings}
\IEEEauthorblockA{\textit{Computer Laboratory} \\
\textit{University of Cambridge}\\
Cambridge, United Kingdon \\
alice.hutchings@cl.cam.ac.uk}
\and
\IEEEauthorblockN{Sergio Pastrana}
\IEEEauthorblockA{\textit{Computer Science and Engineering Department} \\
\textit{Universidad Carlos III de Madrid}\\
Leganes, Spain \\
spastran@inf.uc3m.es}
}
\maketitle
\thispagestyle{firstpage}
\begin{abstract}
In this paper, we describe a new type of online fraud, referred to as `eWhoring' by offenders. This crime script analysis provides an overview of the `eWhoring' business model, drawing on more than 6,500 posts crawled from an online underground forum. This is an unusual fraud type, in that offenders readily share information about how it is committed in a way that is almost prescriptive. There are economic factors at play here, as providing information about how to make money from `eWhoring' can increase the demand for the types of images that enable it to happen. We find that sexualised images are typically stolen and shared online. While some images are shared for free, these can quickly become `saturated', leading to the demand for (and trade in) more exclusive `packs'. These images are then sold to unwitting customers who believe they have paid for a virtual sexual encounter. A variety of online services are used for carrying out this fraud type, including email, video, dating sites, social media, classified advertisements, and payment platforms. This analysis reveals potential interventions that could be applied to each stage of the crime commission process to prevent and disrupt this crime type. 
\end{abstract}


\section{Introduction}
\textit{eWhoring} is the term used by offenders to refer to a social engineering technique where they imitate partners in virtual sexual encounters, asking victims for money in exchange for pictures, videos or even sexual-related conversations (also known as sexting). Packs of multiple images and videos of the people being imitated are traded on underground forums. This material is used as the bait to entice victims into paying for online encounters. Underground forums serve as a place for the interchange of knowledge and new techniques to improve the benefits obtained from this illicit business.

Despite eWhoring being an activity that has been used by offenders for at least eight years, it has received no academic attention. Hence, there is a gap in our understanding about this type of business, how it works, and how offenders profit. By understanding the steps and actions carried out in order to prepare for, undertake and complete such a crime, we can then identify potential intervention approaches \cite{cornish1994a}.

eWhoring came to our attention through our analyses of online underground forums~\cite{Pastrana18a,Pastrana18b}. These communities are used for trading in illicit material and sharing knowledge. The forums support a plethora of cybercrimes, allowing members to learn about and engage in criminal activities such as trading virtual items obtained by illicit means, launching denial of service attacks, or obtaining and using malware. They facilitate a variety of illicit businesses aiming at making easy money \cite{chu:hol,Franklin07,hol,Hutchings15,hutchingsclayton2015,yip:web,zha:tsa}.

In this study we take a qualitative approach, applying content analysis techniques to better understand offenders and their activities. Our data are the discussions between those engaged in eWhoring about how to carry out this fraud. 

The majority of papers in the computer science literature are quantitative, in that they measure incidents and quantify losses. However, by reducing data down to numbers, the inherent richness and meaning can be lost \cite{christie}. There is great value in qualitative research, especially when we need to make sense of the structure and nature of a poorly understood problem. This allows us to work out what to measure in later studies. Thus, while we are not doing quantitative research, dissecting how the eWhoring business is operated is necessary to design further research focused on the different steps (e.g. from the collection of images to monetizing techniques).

The contributions of this paper are:
\begin{enumerate}
\item the provision of an in-depth understanding of the fraudulent eWhoring business model;
\item an applied introduction to crime script analysis, a useful analytical approach used for understanding complex crime types; and
\item a breakdown of the series of steps required to carry out eWhoring, with corresponding intervention approaches.
\end{enumerate}

\section{Background and related work}

\subsection{eWhoring and the law}

eWhoring involves fraudulent behaviour. However, it is not only criminal, but also exploitative, through the deceitful use of images of (usually) young women. eWhoring involves selling photos and videos with sexual content of another person to third parties. This is done by impersonating that person in chat encounters. This misrepresentation to the third party, who pays for what they believe is an online sexual encounter, means this is fraudulent behaviour, similar to romance scams \cite{button2017,huang2015,whitty2012}. 

In this research we find sexual material is distributed in two ways. First, the material is shared with other forum members, either in exchange for money or for free. Second, the material is provided to the customers who are being scammed. In both cases, the distribution is usually without the consent of the person appearing in the photos or videos, or the copyright owner. These images are stolen from various sources, including pornographic sites, social networks, or `revenge porn' sites (which typically contain images that were once shared consensually between partners, but have been leaked online after the relationship sours).

Legal issues may arise relating to the images. For example, they may be indecent images of children. Furthermore, a number of countries have created criminal laws relating to the distribution of `revenge porn' \cite{citron2014}. In the UK under s.33(1)(b) of the \textit{Criminal Justice and Courts Act} of 2015, it is an offence to disclose private sexual photographs and films with intent to cause distress. However, we believe this law might not be applicable to eWhoring, as the \textit{mens rea} element would not be met \cite{henry2016sexual}. The intent of the offender is not to cause distress (someone else has already done that), but rather make a financial gain through fraudulent means. 

We have been unable to find any reference to prosecutions for eWhoring-type activities. This is possibly because victims may be unaware they have been defrauded, or if they have, may be too embarrassed to report this to the police. Furthermore, the limitations faced by police, particularly for low-value frauds, means that they are unlikely to be prioritised for investigation even if they are reported~\cite{wall2007policing}.

In summary, eWhoring could entail a number of criminal or civil offences. Depending on the jurisdiction and actions involved, these could include the redistribution of copyrighted material, the redistribution of material leaked as part of `revenge porn' actions, possession and redistribution of indecent images of children, tax evasion (by not declaring income), and fraud by misrepresentation. As police face limitations when it comes to investigating and prosecuting these types of low-value frauds, understanding and preventing this behaviour from the outset is particularly important. 

\subsection{Crime Script Analysis}

Crime script analysis is an analytical approach used by criminologists and crime scientists to better understand crime problems. Crime scripts break down the commission of crime into a series of steps, from the preparation carried out before it is committed, to after the offence has occurred \cite{cornish1994a}. The universal script developed by Cornish \cite{cornish1994a} has nine standardized script scenes or functions that are arranged in order, namely preparation, entry, pre-condition, instrumental pre-condition, instrumental initiation, instrumental actualization, doing, post-condition and exit scenes. 

While crime scripts are not prescriptive, they provide a useful approach for understanding complex crimes, and hence identifying ways in which they may be disrupted. Crime script analysis borrows from cognitive science, particularly the idea of `schemata', or knowledge structures, that allow us to understand social situations and behave appropriately when responding to others.

Crime script analysis has previously been applied to offences of a sexual nature, including child sex trafficking \cite{brayley2011,savona2013}, sexual offences committed by strangers \cite{beauregard2007}, and child exploitation offences \cite{leclerc2011}. Crime types that are relatively new and emerging due to a strong online component have also been analysed using the crime scripting approach. Examples include the stolen data market \cite{Hutchings15}, the online prescription drug trade \cite{lavorgna2015,leontiadis2015}, credit card fraud \cite{van2016}, fraudulently obtained airline tickets \cite{hutchings2018}, wildlife trafficking \cite{lavorgna2014}, and online piracy \cite{basamanowicz2011}.

The approach taken by Levchenko et al.~\cite{levchenko} when analysing spam advertising pharmaceuticals and counterfeit products also takes an approach that is very similar to crime script analysis. They break down the activities required to successfully market and supply spam-advertised goods, from sending the spam, to payment processing and ultimately completing the transaction. They also identify potential countermeasures designed to disrupt this trade. 

These crime script applications demonstrate the variety of data sources that can be used for analysis. These include surveys or interviews \cite{beauregard2007,hutchings2018,lavorgna2014,lavorgna2015,leclerc2011}, police records \cite{brayley2011,lavorgna2014,lavorgna2015}, court documents \cite{basamanowicz2011,savona2013}, and the infrastructure used by offenders \cite{leontiadis2015,levchenko}. Another data source that is particularly useful for understanding cybercrimes are the forums used by offenders to trade in goods and services, as well as sharing information. Concretely, researchers have analysed written tutorials on these forums, which provide step-by-step instructions for carrying out specialised types of online crimes \cite{van2016,Hutchings15}.

\section{Methodology}

This section presents the dataset of underground forum data (Section~\ref{sect:dataset}) and the method used to automatically extract posts providing tutorials related to eWhoring (Section~\ref{sect:extraction}). We describe our ethical considerations in Section~\ref{sect:ethics}. Finally, we present the  tools used to conduct the crime script analysis in Section~\ref{sect:tools} 

\subsection{Dataset}\label{sect:dataset}
\begin{figure}
\center
\begin{tabular}{c}
\includegraphics[scale=0.25]{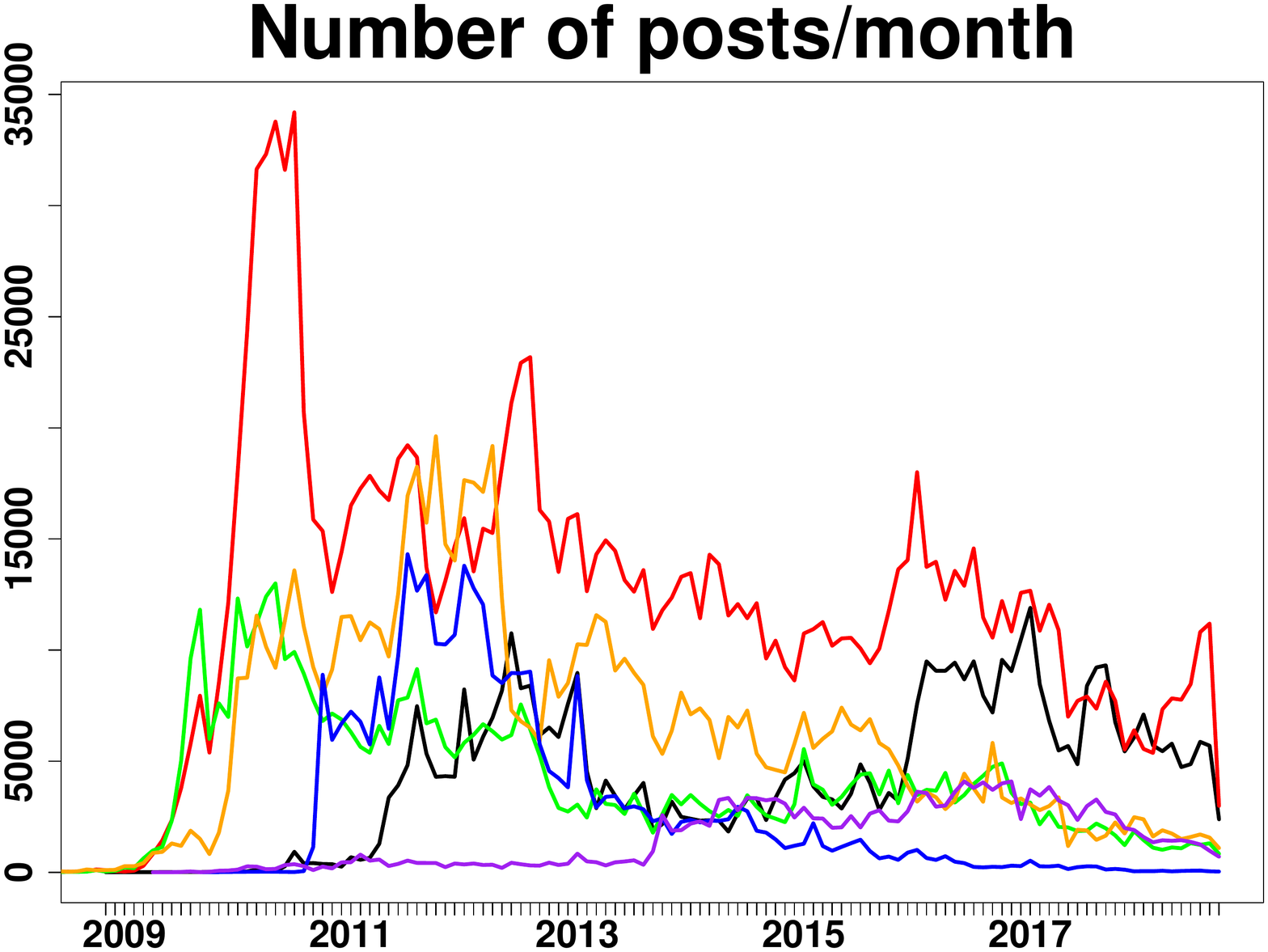} \\
a \\
\includegraphics[scale=0.25]{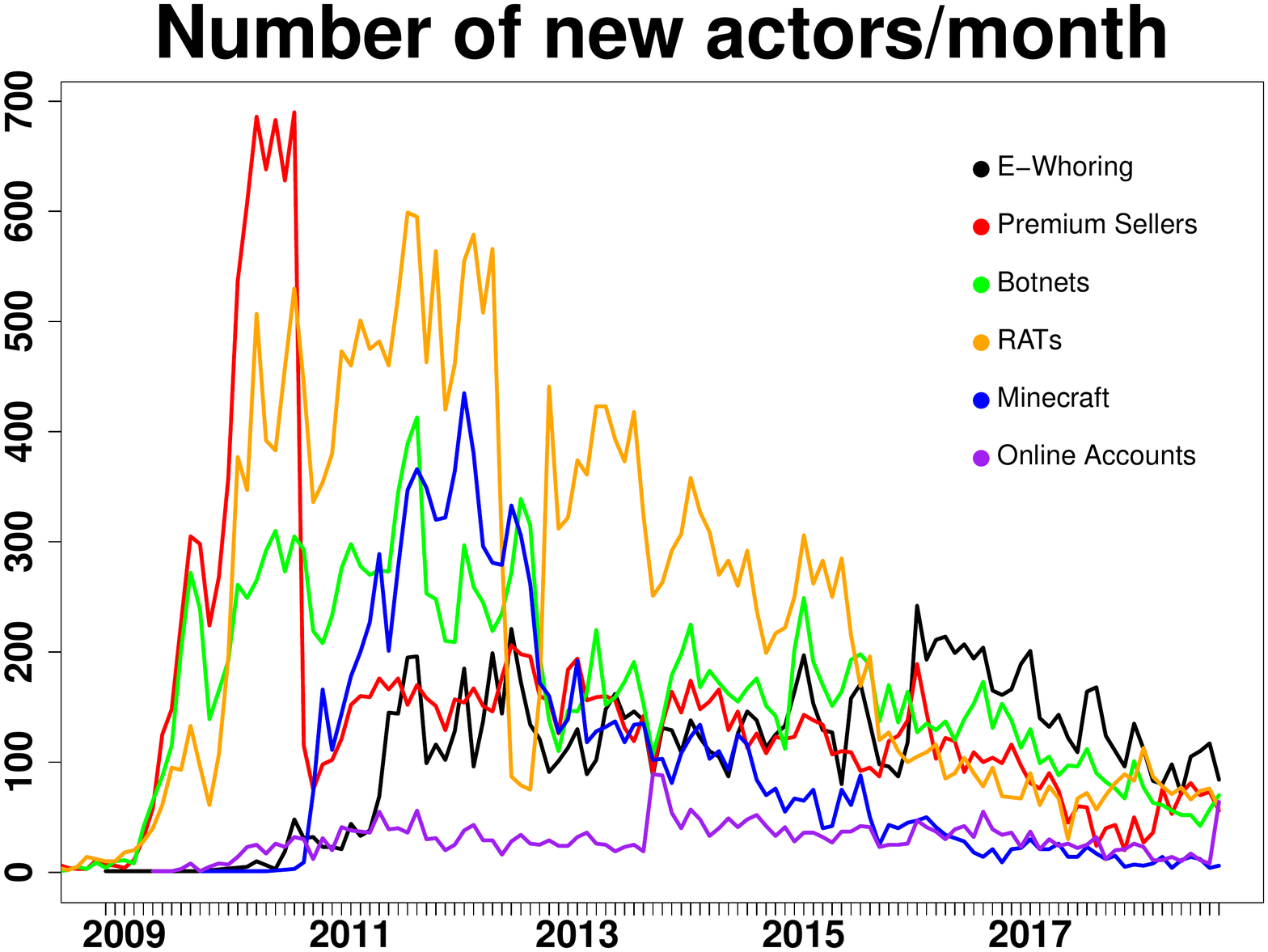} \\
b
\end{tabular}
\caption{Evolution of the number of posts (a) and new actors posting (b) per month, in various popular Hackforum boards}
\label{fig:forumComparison}
\end{figure}

In this work we use the CrimeBB dataset~\cite{Pastrana18a}, which contains data collected from various underground forums. The dataset is available for academic research through the Cambridge Cybercrime Centre.\footnote{https://www.cambridgecybercrime.uk/} We focus our study on Hackforums, the largest forum contained in this dataset, which has a specific section dedicated to eWhoring. Hackforums contains more than 41m posts\footnote{We refer to a whole website as a forum, on which pages are set aside for discussion of defined topics in boards, with users participating in conversation threads via individual posts.} made by 597k user accounts over more than 10 years. Hackforums contains a dedicated board for eWhoring. As can be observed from Figure~\ref{fig:forumComparison}, in the last quarter of 2018 this board has received around 5k new posts per month (at a nearly equal rate to the Premium Sellers Section, a board intended for selling goods and services). Recently, eWhoring has been the board that attracts the highest number of new actors. Around 100 forum users per month write posts for the first time in this board, indicating that this is a popular topic in the underground forum.

\subsection{Extraction of tutorials}\label{sect:extraction}

For this analysis we used heuristics to extract threads that provide guides or tutorials relating to eWhoring. First, we looked for specific words in the thread headings, such as `[TUT]' or `guide'. Then, we filtered out threads that were looking for, rather than providing, tutorials. For this we used a machine learning based classifier to detect whether a thread had begun by asking a question or requesting information (see more details in~\cite{Pastrana18b} and~\cite{Caines18}). Overall, we identified 6,519 posts, written by 2,401 members, in 297 threads, which we extracted for analysis. 

\subsection{Ethical considerations}\label{sect:ethics}
The Computer Laboratory's research ethics committee gave their approval for the research project. Furthermore, we complied with the Cambridge Cybercrime Centre's data sharing agreements. While the data are publicly available (and the forum users are aware of this), it could be used by malicious actors, for example to deanonymize users based on their posts. It was impossible for us to obtain informed consent from users as that would require us to identify them first. In accordance with the British Society of Criminology's Statement of Ethics  \cite{BSCEthics15}, this approach is justified as the dataset is collected from online communities where the data are publicly available, and is used for research on collective behaviour, without aiming to identify particular members. Further precautions taken include not identifying individuals (including not publishing usernames), and presenting results objectively. 

Due to the legal risk of inadvertently coming into possession of indecent images of children, for the research we describe here, none of the images associated with the posts were downloaded. Instead, only text data were collected, excluding all files and images. Standard procedures were also established to enable the researchers to respond appropriately if such material were encountered, namely reporting it the UK's `hotline provider', the Internet Watch Foundation,\footnote{https://www.iwf.org.uk/} which works with service providers to take down child sexual abuse material.

\subsection{Data analysis}\label{sect:tools}

We analysed the forum content using qualitative content analysis procedures. NVivo, a qualitative data analysis program, was used to classify and sort the data. Coding of the data was `data-driven' \cite{gibbs}, in that the categories were selected based on a detailed analysis of the data. As the data coding was completed by one researcher, there were no inter-rater reliability concerns. In addition, a codebook was kept to record the meaning behind each category, and to ensure there was no definitional drift. 

First, a framework was developed using the universal script as put forward by Cornish \cite{cornish1994a}. The various script actions identified through the tutorials were grouped from preparation (scene 1) to exiting (scene 9). This process was iterative, becoming further developed as new information arose in the data. For each part of the universal script we also noted any associated challenges faced by actors, to help inform the potential intervention approaches outlined in Section \ref{sect:interventions}. We also identified a number of `alternative tracks', which are ways in which the universal script has been known to deviate (see Section \ref{sect:alternative}).

The following sections IV--XII outline the script actions for eWhoring. Quotations are provided verbatim for illustrative purposes. On occasion, potentially identifying information has been removed and some quotes have been reduced for reasons of parsimony. Due to the nature of eWhoring, care has also been taken to exclude any explicit content. 

The neutral terms `actor', `customer', and `model' are used below, although these are not necessarily the terms used within the tutorials. We use `actors' to refer to those actively engaging in eWhoring, and/or discussing these activities on Hackforums. `Customers' refers to those purchasing, or potentially purchasing, images. `Models' are those depicted in the images, with or without their consent. 

\section{Scene 1: Preparation}

\subsection*{Learn techniques}

At first glance, eWhoring is not an intuitive business model. Actors have not developed the methods independently. Rather, it is evident that actors have come across eWhoring through their interactions on Hackforums, which is a place for sharing of information and techniques (including the data informing this analysis). The provision of learning opportunities is explicit, through the provision of free tutorials; the advertising of paid `eBooks', which promise to provide more lucrative information; asking and replying to questions; and requests for and the offer of private tuition. 

The forum operates a reputation system, and although not made evident (as this is against the forum rules), it is sometimes implied that tutorials are freely shared with the expectation of receiving positive reputation in exchange. It is also evident that some actors post plagiarised tutorials, copied from elsewhere, presumably to game the reputation system. 

Overall, the tutorials range from being quite general in nature (e.g. providing an overview of basic methods to generate income by eWhoring), to precise step-by-step instructions on how to set up software packages. Some tutorials are specific to certain platforms used to source traffic, such as Craigslist. Others focus on certain methods, such as the use of `VCWs' (an acronym for `video cam whores'), which are described by one actor as:

\begin{quotation}
\noindent A vcw is a program that can control certain videos to smoothly bridge over eachother. For example, your girl is just sitting there, looking around, you can make her wave(If you have the required videos of it ofcourse).
\end{quotation}

\textbf{Challenges.} Actors had to trust the information being provided was correct, amid claims that some tutorials were false or misleading: 

\begin{quotation}
\noindent Come on, good job misleading people AGAIN. Not only you are inexperienced in this section/subject you are making tutorials on it and giving false information and tips. 
\end{quotation}

There were indications that methods providing the most benefit were less likely to be shared with others:

\begin{quotation}
\noindent This one is pretty obvious, if you discover a new site or a new method, be careful who you're telling. You can lose hundreds for that 3rep+.
\end{quotation}

\section{Scene 2: Entry}

\subsection*{Obtain images}

Many tutorials link directly to images that are available on Hackforums, for free and for sale. Images are available as `packs', a selection of photographs, and sometimes videos, depicting the same model. It is deemed preferable to obtain large packs that include explicit images (`nudes'), as well as images that show the model clothed. The latter are used as profile pictures, but also to send as `teasers' to potential customers. Also available are VCWs, which use video footage to enable a customised and interactive `cam show' (the process for making a VCW will be explained below, under `customised images').

Some actors claimed they provided packs and VCWs for free on Hackforums as a `contribution to the community'. However, as found with the provision of free tutorials, there is an unstated expectation that those who appreciate the contribution provide positive reputation in exchange. Furthermore, some actors use a URL shortening service that generates revenue by displaying advertisements to the visitor before sending them to the website containing the packs. 

While the price for packs were rarely mentioned in the tutorials, the going rate appeared to be \$20. For example, the following actor advertised an unsaturated (unlikely to be blocked) `gay' (male) eWhore pack for \$20, payable by Bitcoin:

\begin{quotation}
\noindent I'll give you my unsaturated set (\$20 BTC) which are photos of my friend he doesn't mind ``getting out there''. Includes 200 photos, three verification papers and 4 videos.
\end{quotation}

Advice for making custom packs is readily available. This includes sourcing images from existing websites, such as pornography websites, social media sites (including Facebook, Tumblr, Instagram, and Twitter), and `revenge porn' sites. There were very few references to obtaining images directly from models, although one post discussed socially engineering (here referred to as `SE') someone into voluntarily providing images:

\begin{quotation}
\noindent you can even SE the pictures from a girl if you are having trouble finding a pack (though this is unnecessary hassle since there are hundreds of packs on HF [Hackforums] alone).
\end{quotation}

Three others asserted to have photographs of friends that had been obtained consensually. As shown above, one stated the friend had agreed to have the photos `out there', while the other two claimed their friend had posed in exchange for money. One actor claimed to use recorded `live shows' they had purchased from pornographic websites, which allowed them to request certain poses for verification purposes:

\begin{quotation}
\noindent Instruct your model not to move too much while doing stuff like ``wave'', ``peace sign'' etc. This will make creating the VCW so much easier. [This adult site] is a great site to get models. Ask for a private show AND ENSURE THE MODEL ENABLES THE RECORDING FEATURE. You will then find the recording in your buyers collection. 
\end{quotation}

Verification is a topic of concern on Hackforums. Some customers are suspicious and request proof, such as the model being in a certain pose. This explains why the actor above seeks footage containing certain gestures. In relation to still photographs, different types of verification may be requested. For this, a `verification template' is useful. This is an image with surface area on the body or piece of paper that can be easily altered to include the customer's name. One actor provided a tutorial for obtaining verification templates that capitalises on a social media craze that began in 2016, the \#A4challenge. Social media users posted images of themselves holding a blank piece of A4 paper, to demonstrate the size of their waist. Images of people of a similar appearance to models in the actor's packs are apparently useful for verification purposes:

\begin{quotation}
\noindent Open up the instagram app. Search for the hashtag \#a4challenge [...] There are loads of images here that you could use, I'd advise that you go through and save a number of them, that way you can make sure it fits the particular eWhore pack your using at the time. (The one i use at the moment is from that list, its one of the faceless ones).
\end{quotation}

Overall, low quality, `amateur' images are preferable, as `professional' images and videos can invoke suspicion. This is also useful for creating VCWs, to disguise the transition between poses:

\begin{quotation}
\noindent Videos are low quality and laggy but that's exactly what makes it so effective because the transitions aren't too obvious.
\end{quotation}

\textbf{Challenges.} One challenge was obtaining images that were not `saturated'. Using the same images as others may not only be problematic when engaging with potential customers (who may try a reverse image search), but also for ensuring advertisements to attract customers are not flagged and taken down:

\begin{quotation}
\noindent Apparently if you upload an ewhore pic, it automatically blacklists you. I signed up \& as soon as I got to uploading the pic, I used one of my favorites \& Bam, I get logged out and it will not let me get back in, sayin I don't have an account. When I try to recreate it with the same email it says I already have one.
\end{quotation}

There are also risks associated with the files themselves. If the images are of underage models (this was generally not recommended), there is an obvious legal risk. Furthermore, there were concerns that accessing some of the packs might lead to malware infection.

\section{Scene 3: Pre-condition}

\subsection*{Create an alias and prepare a backstory}
Once images have been obtained, a fictitious identity is created. This begins by bestowing a name on the model depicted in the images. Some tutorials went to the extent of suggesting actors use an online name generator to pick a suitable alias:

\begin{quotation}
\noindent Now that you have some pictures you need a total new identity. Please Click Here [http://www.fakenamegenerator.com] And randomly generate a fake FEMALE name. Right it down [...]
\end{quotation}

A number of backstories were provided as examples. These ranged in complexity, from demographic information to explaining why the model needed money:

\begin{quotation}
\noindent To start ewhoring, you'll need an identity, that means; [*]A name [*]Where you are from [*]How old you are [*]Why do you need the money? [*]Why not working? [*]Family? [*]Studying? College? These are the most vital things you'll need to know. ALWAYS use the same info, as some people tend to check you with a second account. Get it fixed in your brain before starting out.
\end{quotation}

Not only should backstories be consistent, they should also be unique to each pack. Some backstories were designed to socially engineer the customer into paying more, for example:

\begin{quotation}
\noindent Your target will more then likely ask you , ``how are you''? You can say something like, ``not so good :('' [...] You can say that you've been kicked out your house and you have no money. If he asks where are you just now? Say you're living with a friend, but can't live there very long. You can also say you need help paying rent. -- Another good method is to say you're little sister or brothers birthday is soon, you're too broke to buy anything for him/her and would do anything for some spare cash.
\end{quotation}

\subsection*{Open accounts }
The alias is used to open online accounts that will be used for subsequent steps. Accounts are typically required on communication platforms, payment platforms, and websites used to attract traffic.

An email account is one of the first account types that is required. As will be discussed later, in some cases an email account will be used to correspond with customers and to receive payments. However, an email address is often also a prerequisite for opening other types of accounts. Suggested webmail providers included Yahoo!, Gmail, Hotmail (when it went by that name), and AOL Mail. Other communication platforms include Skype, Kik, Snapchat, Facebook, and even text message (SMS). Some of the platforms recommended have since been discontinued, notably AOL Instant Messenger and MSN Messenger. Some, such as Kik and Snapchat, are primarily mobile phone applications, so actors can choose to just use their phones, or install a phone emulator on their PC. 

There are two primary platforms recommended for receiving payment: PayPal and Amazon (through gift cards). A couple of actors suggested receiving payments through Western Union, and one tutorial from 2012 suggested `Liberty Reverse', which is assumed to mean Liberty Reserve, a digital currency provider that was shut down in 2013. For PayPal, there was some discussion about whether to create a `verified' or an `unverified' account. Verified accounts are linked to a bank account or credit card, and therefore require it to link to a real identity:

\begin{quotation}
\noindent Login into PayPal and click ``Add or Remove Email'' Then, add the email that you're using to ewhore as a secondary email. What this will allow you to do is to receive money to both your normal email and you ewhore email. This will remove the need to have two accounts and the difficulty of getting money from one account to another. However, this does have it's drawbacks. When money is being sent to you, from either email, it will show the your name. To work around this I say that the money is for my rent and me and my brother/partner/roomate shre one PayPal and the rent comes out of the attatched account. 
\end{quotation}

While unverified accounts can be created using the alias, the actors suggest they are more likely to be flagged as suspicious, and there are restrictions on making transfers. One innovative suggestion to receive payments through a third-party site is to set up an account on a freelancer marketplace:

\begin{quotation}
\noindent Make a fake account on Fiverr, but make sure to use a general girly name because you are going to use this account for all your ewhores. Think of some really easy to do gigs, something like ``make a Photoshop edit''. [...] Tell your pedos\footnote{Some actors referred to potential customers as `pedos'. In most instances it did not appear there were accusations of or links with pedophilia, but rather the actors were using terminology to imply the customers were somehow deviant.} that you want to keep your identity secret by not using your real PP, or that you want your family to think that you're doing some decent job on the internet and not sell nude selfshots or videos. So, tell the pedos to buy your gig a couple of times until you reach the desired amount of money [...] the money will go straight into your real PP account. 
\end{quotation}

Many tutorials also recommended the use of Virtual Private Networks (VPNs) or proxies, and there were suggestions about the best services to use. However, there was debate about when an IP address should be hidden. Some did not feel the need to use a VPN or proxy at all, while others recommended use at all times. Some tutorials recommended actors hide their IP address only for certain actions. For example, one actor recommended not hiding IP addresses when opening accounts on traffic sources, but to do so when communicating with customers:

\begin{quotation}
\noindent NEVER use a proxy when posting your ads on CL [Craigslist], when i first started this non of my ads were getting posted so i thought screw this and did it without a proxy and it worked, so never use a proxy if you want to your ads to post properly. Always use a proxy when your emailing back to the clients/buyers, people can trace you through your emails which you have sent and will be able to find out where your from. This is uncommon but it is heard off.
\end{quotation}

Another actor also recommended not using a VPN when opening accounts, but suggested that when an account was banned, users reset their (dynamic) IP address before creating a new one:

\begin{quotation}
\noindent Never use a VPN, [...] just stick a pen in your router to reset it or use command prompt etc; every time you post a new add and every time you make a new email to get a new IP. 
\end{quotation}

\subsection*{Customise images (optional)}

There were only a few, rather vague, tutorials for creating VCWs from video footage. The following is an excerpt from one of the more useful tutorials, although this does not use Adobe Flash Player, which was mainly referred to elsewhere (instead it uses a webservice that, at the time of writing did not appear to be available):

\begin{quotation}
\noindent Timecuts is intended to cut to different scenes in video footage and can be used to make VCWs easily. [...] The free version has basic stuff but the premium one enables you to stream your VCW as a ``fake webcam'' to Skype, Firefox, Chrome etc. without need for screen recorders and other features that make your VCW more convincing like fluid loops. When you open TimeCuts, there is an ``Add Scenes'' tab, a ``Transitions'' tab and a ``Run'' tab. Add Scenes tab [...] Add all the scenes you need like ``wave'', ``smile'', ``stand up'' etc If your model doesn't move their body around much, this is pretty much all you'll need. The premium version has ``Fluid loops'' which essentially enable you to loop forward and backwards on the same clip making it look continuous. Transitions tab [...] You can use this tab if you need to add small clips to make a transition between any two scenes [...] for example if your model is chat looping with their shirt pulled up, you'll want them to pull it back down before the next action that is, if the next action they're doing is with their shirt pulled down. Otherwise it wont look right for them to go from no shirt to shirt in a second. Run tab [...] It will automatically run the main loop you specified such as ``typing'' then you can easily switch between scenes by clicking the various buttons.
\end{quotation}

Other tutorials provided instructions for preparing verification templates, removing watermarks from images, changing the date and time a photo was taken, changing images so their source cannot be identified through reverse image searches, and adjusting the brightness:

\begin{quotation}
\noindent Sometimes you Ewhore at Night but your VCW is showing of in a bright room ? Just Adjust the brightness to make it much more realistic. 
\end{quotation}

\textbf{Challenges.} Verifying accounts, such as providing telephone numbers or bank details, was problematic for many actors. In some cases, this was seen as an operational security issue, due to concerns about the legality of their actions:

\begin{quotation}
\noindent Paypal and amazon ask for credit card info, but it's not like I have a fake credit card lying around. If I use my own personal credit card, but make the amazon account with a fake email, can it ever be traced back to who I am? 
\end{quotation}

In other cases, it was more of a practical matter. For example, there were many discussions about what to do when Craigslist requested telephone verification when setting up an account, as the same number could presumably not be used across banned accounts without also tainting new advertisements.

Customising images seems to be technically challenging for some, and tutorials for operating VCWs were often followed by a large number of posts with people complaining of difficulties (e.g. `I'm about to give up... been trying this for a long time now but i keep getting stuck at the same problem').

\section{Scene 4: Instrumental pre-condition}

\subsection*{Source traffic}

The next step involves meeting and responding to the initial contact from potential customers. This requires creating an advertisement or profile on the platform being used to source traffic. The following example is taken from a tutorial about eWhoring on Craigslist:\footnote{We note that in April 2018 Craiglist removed the Personal section on its US site in response to the US FOSTA/SESTA bill, which makes websites liable for advertisements that promote or facilitate online sex trafficking.} 

\begin{quotation}
\noindent It is now time for you to post your classified advertisement. You must start by selecting the Personals category. [...] When posting, make sure to have an appealing title. Something along the lines of ``Hot 18yr LOOKING FOR OLDER MAN'' should work perfectly. Don't forget to add some pictures [...] It is probably a good idea to not display any nudes. You should save those for the emails. Once your advertisement is posted, you should begin to receive emails. 
\end{quotation}

There were 176 unique websites and applications suggested as traffic sources. These could be broadly classified as: chatrooms (n=120), many of which were `adult' themed; video chat, such as Omegle and Chatroulette (n=12); social networks, such as Facebook, Instagram and Tumblr (n=10); amateur pornography (n=4); classifieds, such as Craigslist (n=7); dating, including PlentyOfFish, Grindr and Tinder (n=21); and online gaming (n=2). However, it was noted many of the sites listed on Hackforums were `saturated', and actors were likely to keep particularly useful sites to themselves:

\begin{quotation}
\noindent Traffic is the only thing that will get the buyers to you. You have 2 types of traffic, saturated and unsaturated traffic. Saturated traffic can be found everywhere, they will get you buyers, because there will always be new people joining them who don't have a clue how things work on kik or those apps/sites. The unsaturated traffic won't be easily found. Nobody on HF will give our there source of HQ traffic. You'll need to find it for yourself. It's most of the time a goldmine, but you'll probably need a custom pack for it, but it's worth it.
\end{quotation}

Some tutorials suggested openly advertising photographs and cam shows for sale (for example, `in need of cash asap selling my p1ctur3s and maybe more. Pm for details xx'). Others recommended broaching the subject after communication had been initiated. The following example is attracting traffic from a chatroom:

\begin{quotation}
\noindent So to really start off you can go to any chat website the easiest you can start with [is Chat-Avenue], say something like this ``Hey guys! Wan't to add me on skype to chat? YOURSKYPE and then you can watch all the pervs adding you on skype, then you just put your social skills to work and you should make progress in the first hour or two.
\end{quotation}

As shown above, this step often includes transferring potential customers to a separate communication platform. This is to avoid being banned from the traffic source (`no dating site wants to become a happy place for E-Whores'). The choice of communication platform should correspond with the type of images to be traded, for example, `Put e-mails if you want to sell pics, put skype if you want to sell shows'.

This is the first step that involves interaction between an actor and a potential customer. Advice was plentiful, albeit often stereotypical, on how to be flirtatious and come across as believable, for example:

\begin{quotation}
\noindent Making sure to spell incorrectly and using emojis. The main purpose of this is to make sure you seem like a legit girl. Girls who text never use the perfect grammar and they always use emojis at the end try to prevent from the x's at the end because that is very over used. The point I am trying to get across is text like a girl.
\end{quotation}

Some actors recommended developing macros and other ways of automating the communication, to avoid repeatedly typing similar phrases:

\begin{quotation}
\noindent For example, shortcuts on your keyboard, can be used on pc and phone [...] For example on the phone, if you type ``mes1'', it will autocorrect to [...] ``Hey there, how are you? Interested in buying some pics? ;)''
\end{quotation}

Tutorials contained a number of recommendations, such as which cities to post Craigslist advertisements in, and which days and times corresponded with more requests.

\textbf{Challenges.} One of the main challenges for sourcing traffic is advertisements being flagged and taken down, and accounts closed. While some were reported by other users, it was also apparent that platforms are automating the detection of suspicious accounts: `Ummm, well I got banned in about 13 seconds. Facepalm ;('. Actors continually had to adapt their advertisements to avoid them being taken down. This also seems to be a reason why few posed as juveniles, as these were more likely to be flagged (in addition to the legal risks associated with distributing images of children).

Another challenge related to interacting with potential customers. Those that did not want to pay were referred to as `timewasters'. The lack of immediate gratification was a disincentive for new actors to continue:

\begin{quotation}
\noindent I've been ewhoring for 2 whole days and havent been able to earn a cent. [...] Im about to give up on ewhoring 
\end{quotation}

Some of the actors indicated they were recipients of misogynistic abuse (e.g. `they will turn around and call you a whore and every other name in the book'). While this did not seem to especially resonate with them (in fact, there was content within discussions on the forums that could be described as misogynistic, as well as homophobic, ableist, sexist, and racist), there were concerns about receiving unsolicited explicit images from potential customers:

\begin{quotation}
\noindent -DONT LOOK AT THE SHIT THEY SEND YOU (all [explicit] basically... will scar younger people for life lol)
\end{quotation}

\section{Scene 5: Instrumental initiation}

\subsection*{Negotiate}
Actors were proud of their social engineering skills, and bragged about their abilities to turn difficult conversations into profitable situations. This mastery over others seemed in itself to be perceived as a benefit:

\begin{quotation}
\noindent This is an art. The art of manipulating people in your favor. Getting out of every tight sittuation. Coming up with straight up answers on things, imagination. This can't be tought, this must be learned.
\end{quotation}

Many actors mentioned that potential customers would ask for `previews'. While some admitted sending nude previews, this was discouraged, and it was instead recommended that images of the model clothed, or partially clothed, be sent instead:

\begin{quotation}
\noindent They most likely will ask for a preview before buying, don't send a nude as stated above. Send a pic that shows alot of skin, [nothing explicit]. Cover them up, send one in a sexy bikini or covering everything with your arm, and make them want you.
\end{quotation}

Another frequently discussed request was verification images, such as the model displaying the customer's name written on their body or a piece of paper. Obtaining verification templates has already been discussed as an earlier step. Tutorials for customising these to the customer's requirements in a realistic way were provided. Previews and verification were also requested for cam shows:

\begin{quotation}
\noindent Ask them if they want you to wave on cam [...] If they ask you to verify you're real by showing [explicit content] DO NOT DO THIS, be hard! Ask them to verify they can pay [...] If they ask you to stand up and your VCW doesn't have that option, tell them you are busy finding people who possibly want to buy a show and can't stand up whenever they want you to. 
\end{quotation}

The going rate for selling pictures and cam shows was contentious. As one actor said, `Since e-whoring is 100\% profit, you can pick any price you desire'. However, others firmly believed that selling too cheaply hurt the eWhoring community overall:

\begin{quotation}
\noindent That isn't even realistic + way too cheap.. Make it \$20 for 50 or whatever.. Otherwise the other eWhorers will look expensive as fuck and you will ruin it for the others...
\end{quotation}

While some actors had a price list for their products (for example, `My usual rates are: 10\$ for 15 pics, 15\$ for 20 pics, and 20\$ for 25 pics'), others let the customer suggest a starting price and then negotiated from there. 

\textbf{Challenges.} The main challenge here is being able to negotiate a sale, providing verification, and dealing with `freeloaders', who tried to get free previews, but would refuse to pay:

\begin{quotation}
\noindent These freeloaders will add you on skype and ask for pictures. I reccomend giving 1 for free [...] They will try and get as much out of you for free then block you.
\end{quotation}

\section{Scene 6: Instrumental actualization}

\subsection*{Receive payment}

Once a price has been agreed, but before the images are sent, payment is received. As mentioned, PayPal and Amazon gift cards are the most commonly accepted payment methods, and both allow transactions to be sent using just the recipient's email address. Tutorials instructed actors to request PayPal payments using the `gift' or `donation' options, as this apparently prevents customers from later requesting a chargeback.

Before proceeding to the next stage, the actor should verify they have received payment, to avoid being scammed by the customer:

\begin{quotation}
\noindent Don't send anything before you've checked your balance. Never. They're [...] pretending they have paid, by showing you a screenshot of a fake email from paypal or amazon. Don't believe it untill you see it for yourself.
\end{quotation}

\textbf{Challenges.} Some actors reported problems with customers requesting chargebacks after the transaction was complete. This was a particular problem for payments received using PayPal.

\section{Scene 7: Doing}

\subsection*{Send images}

Photographs can be sent using the communication platform, or by sending a link, for example:

\begin{quotation}
\noindent Then proceed to asking how he would like to get the pics. I most of the time send them a link to a dropboxlink or just sending the pics to his email. Easy peasy! 
\end{quotation}

Cam shows are typically delivered with the assistance of an application such as ManyCam. This application is used to trick the customer into believing they are viewing a live camera feed, while the actor controls the video footage. The video settings in the communication platform are adjusted to select ManyCam, rather than an internal or external camera. Many tutorials advised how to use and configure ManyCam, such as:

\begin{quotation}
\noindent [...] open up many cam you should get something that looks like this [screenshot]. Now go ahead and click the movies tab on the left. [...] Click add files and select your movie clip [...] Now to get this camera to show in Skype [...] under ``call'' go to ``video'' then ``video settings'' and there should be a drop down box that has your selected webcam. Change that webcam to ManyCams webcam and your done. 
\end{quotation}

\textbf{Challenges.} Cam shows can be difficult to coordinate successfully, such as ensuring the transitions do not appear unrealistic. Actors also recommended ensuring audio was turned off: `Also make sure your audio is NOT enabled, you don't want these guys hearing you laugh at them'. 

Some actors found their moral beliefs was a disincentive to continue (`Though it was great money, I just didn't feel morally right doing it').

\section{Scene 8: Post-condition}

\subsection*{Block or continue to milk customers}

Some actors recommend ceasing all contact with customers after the first transaction:

\begin{quotation}
\noindent When the trade is done, you might want to leave the chat (unless he asks for more) or block him from your msn. On to our next target!
\end{quotation}

However, others suggested initial customers could be a source of future income:

\begin{quotation}
\noindent once you have receive your payment send him the pics, and do not delete or block him straight after as he's a possible source of future incomes, carry on talking to him and make him spend more, say things like ``i saw a really nice vibrator on amazon the other day and i though of you maybe i could give you a show ;) xx''.
\end{quotation}

There was also a suggestion that a good customer could be targeted using more than one model:

\begin{quotation}
\noindent Milk all your customers as much as possible. Get every last dime out of them, without scamming ofcrouse. And when your done, start over with a new account and message that same guy, and milk him again!
\end{quotation}

Some actors suggested continued discussion with customers to scam them for money, such as for flights so they could visit, similar to romance scams. For example:

\begin{quotation}
\noindent for the next 2-3 weeks [...] just talk to them, tell them about what you've (supposedly) been doing that day, [...] most importantly, appear interested in them, appear to genuinely like them. [...] when you know each other a little [...] act really worried, just be like ``help, I have a really big problem, my dad found out about the pics I sent you, now I have to go stay with my friend for a few days'' - keep swearing, not directly at them, just at life in general, ``fuck, I don't know what to do..'', ``shit.. I'm getting kicked out and have nowhere to stay'', now's where you throw out the real trap, say something like ``I know you probably dont have room but.. is there any way I could stay with you for a few weeks? I'd let you see me naked and.. maybe a little more I guess'' [...] when they agree [...] say ``fuck how am I gonna get to you? 1 sec I'll check flights'', look up the price of flights, then get really upset. ``Oh god, it's going to cost me \$900 to get there, and I only have \$135, what am I gonna do? )))):''
\end{quotation}

\textbf{Challenges.} Some actors were concerned they would be found out, however the likelihood and consequences of this risk were not seen as severe:

\begin{quotation}
\noindent In some cases you will be caught and the guy will know you are fony [...] He really can't do much against you so no worries :)
\end{quotation}

\section{Scene 9: Exit}

\subsection*{Exchange funds}

The last step is to retrieve the funds from the account they have been deposited into. If using PayPal, accounts may be verified or unverified. Funds deposited into verified accounts can be transferred to a linked bank account. Unverified accounts are trickier, as they have limits on the amounts that can be sent, and are more likely to be flagged for suspicious activity. There were a number of suggestions for handling unverified accounts, which differed according to the country the account had been set up in.

Amazon gift cards appeared less risky than PayPal. There was less risk in disclosing the actor's identity, and no complaints relating to chargebacks or account limitations. Some actors used them directly to purchase goods, `I love using amazon: ) I just buy a lot of things off of there.. \textless3'. For those that preferred to spend their money in other ways, instructions were provided for converting Amazon gift cards to Bitcoin using PurseIO, although commentators argued the conversion rate was low. Other sites to convert Amazon gift cards included reddit.com/r/giftcardexchange/ and the currency exchange boards on Hackforums. 

\textbf{Challenges.} PayPal and Amazon gift cards each had their own challenges. For PayPal, these were avoiding chargebacks and having the account limited for suspected fraud, suspicious activity or abusing their terms of service. However, the difficulty with Amazon gift cards is being able to monetise them without losing too much of their value. 

\section{Alternative tracks}\label{sect:alternative}

Alternative ways of generating income from eWhoring were also outlined. From blackmail to double dipping, these are listed below in order of how often they were discussed. While some actors were proponents of such approaches, others recommended not doing these activities, as they introduce additional risks to the individual (for example, blackmail), or to the eWhoring trade as a whole (such as scamming customers). 

\subsection{Blackmail}

Despite discussions relating to blackmail apparently being banned on Hackforums, there were nine tutorials including suggestions for blackmailing customers after they had purchased images. For example: 

\begin{quotation}
\noindent Unfortunately there are sick pedophiles in the world. [...] We can play this to our advantage if you don't mind resorting to blackmail. This can easily bring in \$3,000 a week. However, the dangers are higher. [...] Go into a chatroom and say something like ``17 year old looking for an older guy to talk to''. You will get bombarded with replies. [...] Leave the chatroom, but have him add you on MSN or email you. [...] Then tell him [...] you didn't want to scare him off, but you're only 16. If he freaks out and won't talk to you, move on to a new person. If he is okay with it, keep going. [...] Write down everything you learn about him. His age and location, his email address, everything. [...] A few days in, I'd be surprised if he didn't ask you for a nude. [...] If not, make him want one. [...] Without him willingly taking a naked picture of what he thinks is a minor, all of your work was pointless. [...] Email your pedo scaring him shitless. Say something like ``Hello Mr. (Pedo's name). What you have been doing is illegal. You willingly took a picture of me, a minor. I have chat logs and screenshots. You will go to jail for a long, long time if I tell anyone. I bet (spouse's name) would be mad. What would the people at (workplace) think? That speeding ticket you got years ago (scaring him so he knows you know everything about him) would be meaningless compared to this. I expect a monetary compensation of (your choice) in my Paypal by (your choice of time). If you block me or stop communication with me, I will call the authorities to (Pedo's address here).'' Sound professional. Don't copy mine word-for-word. Make sure he knows you are serious. 
\end{quotation}

While all the tutorial scenarios related to the alleged age of the model, and the legal implications, it was noted there was an adultery angle that can be capitalised (`Not always pedophilia though. A legit online fling with another grown women'). 

There were acknowledgements of the serious legal risks relating to blackmailing, and the need to take stringent steps to ensure anonymity. However, there were also indications actors felt it was morally just to blackmail those who would knowingly exploit children.

\subsection{Affiliate marketing}

Affiliate marketing involves diverting traffic to certain sites, for a fee per click, or a percentage of subsequent sales. Tutorials mentioned a number of ways actors could incorporate affiliate marketing into their activities. One way is to use URL shorteners that first display advertisements to the visitor. Other links require the visitor to complete a survey in order to view the contents. Another method attempts to monetise potential customers that are unwilling to pay for images, but offering them free photographs in exchange for them downloading a mobile application:

\begin{quotation}
\noindent This website will pay you each time you are able to successfully get someone to download FREE apps on their iphone/android. Here is what it will look like for your clients when you send them your link: [screenshot] If you successfully get [timewasters] to complete the offer, you will receive around 40 cents per installed application [...] add that ontop of your e-whore earnings, you can profit an extra 50-60 dollars daily. 
\end{quotation}

One actor mentioned they made `\$15 for every user that does an ``age verification'''. This involved sending a potential customer to a website to enter their credit card information under the guise of ensuring they are not a juvenile, while making false promises they will not actually be charged anything. 

\subsection{Send malware}

Another way to benefit from eWhoring is to spread malware to potential customers, by attaching the malware to the images or having them visit a website with a drive-by download. There are a number of ways malware can be monetised, such as keylogging credentials to gain access to online accounts. In the following example, the intention is to enslave the customer's computer into a botnet:

\begin{quotation}
\noindent get a good crypter, bind a RAT [remote access trojan] to the noodz, and send em away, for free OR money. then easy slave for my botnet ;)
\end{quotation}

\subsection{Scams}

Scamming customers is the easiest alternative track, and is also much easier than eWhoring itself. Rather than including additional components into the eWhoring script, it omits the steps relating to obtaining or sending images. A number of tutorials were available that recommended scamming. However, some actors had a moral objection to this practice:

\begin{quotation}
\noindent Don't scam, just don't. You wouldn't like to get your money stolen in a store either. Provide what they paid for, so everyone is happy.
\end{quotation}

There are some other objections to scamming, perhaps from those that are selling packs (who have their own incentives to promote eWhoring). The reason given for these objections is that scamming makes it harder for others to entice customers:

\begin{quotation}
\noindent Always deliver to your slaves. People who scam slaves do bad for all other eWhorers and themselves.
\end{quotation}

\subsection{Double dipping}

This alternative track requires an amendment to the `receive payment' stage. The customer is tricked into thinking there has been a problem with their initial transaction, and are persuaded to send the money again. Tutorials recommended amending the payment receipt to say payment was not received, and sending a screenshot to the customer:

\begin{quotation}
\noindent The double dip method. This is a very useful method because it will [...] double the money [...] highlight Payment received and right click and click on inspect element [...] make a excuse up saying for example ``Payment error'' or ``Payment not received'' [...] screenshot it and send it over to your victim. And hopefully they will send you double the amount they already sent you. This can be done the same way on a amazon gift card all you have to do is do the steps for PayPal that exact way.
\end{quotation}

\section{Intervention approaches}\label{sect:interventions}

\begin{table*}
\caption{Potential intervention approaches for eWhoring}
\begin{center}
\begin{tabular}{l l }
\hline
\textbf{Script stage} & \textbf{Potential intervention} \\
\hline
Preparation: Learn techniques & Promote distrust relating to tutorials \\
\hline
Entry: Obtain images &Increase likelihood of image saturation \\
\hline
Pre-condition: Create an alias, prepare a backstory, and open accounts & Verify accounts and shut down for misuse \\
\hline
Instrumental pre-condition: Source traffic & Detect advertisements \\ 
\hline
Instrumental initiation: Negotiate & Demand reduction \\
\hline
Instrumental actualization: Receive payment & Shut down payment accounts \\
\hline
Doing: Send images & Detect and block saturated images  \\
\hline
Post-condition: Block or continue to milk customers & Anonymous reporting \\
\hline
Exit: Exchange funds & Regulate the exchange of alternative currencies \\
\hline
\end{tabular}
\label{tab:prevention}
\end{center}
\end{table*}

Crime script analysis is particularly useful for identifying new ways to intervene and disrupt crime. Table~\ref{tab:prevention} provides an overview of potential intervention approaches that could be applied to each point in the script. We describe these potential approaches, discussing their feasibility and their potential impacts on the law-abiding majority. We also note where emerging technologies may provide additional opportunities for eWhoring, leading to future innovation. 

\subsection{Disrupt tutorials}

As eWhoring is not an obvious crime, one way to disrupt it is to remove the tutorials, which are concentrated on Hackforums. While they may reappear in other locations, it is likely to be more fragmented. This would make it more difficult for engaging new actors. Additionally, the entire section dedicated to eWhoring could be removed. Technically, this will require the cooperation of forum administrators. There is a similar precedent: an entire section intended for the trading of \textit{booter} services was removed, after a user released the code of the botnet Mirai, used to launch several DDoS attacks~\cite{KrebsHFShuttersBootersBazar}.

Another method could be to promote distrust relating to tutorials. There are already some indications of untrustworthy behaviour, such as plagiarising tutorials. This is presumably an attempt to game the reputation system, in the hope that others will bestow positive reputation to thank them for their contributions. This may be of concern to the forum administrators, and they might want to introduce anti-plagiarism checks to avoid this form of gaming. 

Another approach that would not require the intervention of the forum administrators is to create `slander' accounts \cite{Franklin07}, particularly about reputation gaming. Slander accounts could also be used to question the veracity of tutorials, and promote concern about the risks involved in eWhoring. This approach would require the creation of accounts and communication with potential offenders. This can raise ethical issues, and the manual effort required would be time consuming.

\subsection{Increase likelihood of image saturation}

One challenge for obtaining images is to find ones that are not `saturated', or blocked on the sites where they are to be placed. To enhance the likelihood an image will be saturated, or that known images will be detected on sites, a hashlist of known eWhoring images could be developed. Similar to hashlists of child sexual abuse material, known hashes could be added to the list at the time of detection. The hashlist could be shared between social media and dating sites, hosting providers, and communication platforms. They could then use this hashlist to automatically detect and block eWhoring activity.

There are some technical and operational drawbacks with this approach. The main limitation is images can be intentionally modified to change their hash value while keeping the sexual content, e.g. by mirroring the image, or changing its brightness. Thus, the hash algorithm would need to consider all such possible modifications.

\subsection{Verify accounts and shut down for misuse}

Some service providers that are misused for eWhoring frustrate actors by requiring account verification. In particular, telephone verification is a barrier for some users. While blackmarket services may offer to get around verification requirements, this adds an additional layer of cost and complexity to the crime commission process. Therefore, platforms that find themselves being misused for eWhoring may consider implementing a verification process, and not allowing multiple accounts be created with the same verification details within a short period of time. Furthermore, these platforms may implement systems designed to detect characteristics indicating eWhoring activity, and shut down the accounts and blacklist the users if they are found to violate their terms of service.

The feasibility of this intervention depends on the incentives for implementation. Particularly when the costs of eWhoring are not felt by the service provider, the time and costs required to implement verification may be seen as excessive. Furthermore, some providers will be concerned about the impacts on legitimate users, and the ease with accounts can be set up on their platforms.

\subsection{Detect advertisements}

Actors source potential customers by advertising or initiating conversations in chatrooms, video chat, social networks, amateur pornography sites, classifieds, dating and gaming sites. An effective intervention which frustrated actors was detecting their advertisements, shutting down their accounts, and preventing the opening of new accounts, which some platforms appeared to be better at than others. Automated systems may be effective in detecting suspicious advertisements \cite{stringhini2010} to be targeted for takedown. It is important that service providers understand how these accounts are being monetised, as otherwise they may consider them to be a nuisance, rather than criminal.

A derived problem of this approach is the potential harm for legitimate sex workers, whose advertisements could be unfairly banned. Establishing the limits on which advertisements are from real models and which are not is challenging. Indeed, eWhoring actors could mimic ads used by real models, similarly to what they do during the actual online sexual encounters with customers.

\subsection{Demand reduction}

One way to reduce demand is to increase awareness among potential customers. `Honey' accounts, mimicking eWhoring accounts, could be set up. Instead of exchanging images, potential customers could be provided with information about eWhoring. The message could focus on: the dangers customers potentially face being further scammed, blackmailed or infected with malware; they may not be talking with the person they think they are; and how the images of young women are being exploited. One problem with this approach is identifying those with the incentive to dedicate time and resources to implementing the initiative. Another less time intensive (but still costly) approach may be educational campaigns, with advertising on the sites used by actors for eWhoring. 

\subsection{Shut down payment accounts}

It was evident PayPal has been making efforts to reduce the use of their accounts for fraudulent purposes. This had led to a subsequent move to Amazon gift cards. This type of displacement has been seen with other crime types, for example \cite{karami2016} evaluated the effects of PayPal shutting down the accounts linked with the provision of denial of service attacks, and found operators displaced to Bitcoin. Despite moving payment platforms, the intervention still had a negative effect on their profits. 

However, despite the efforts made by PayPal to close fraudulent accounts, it is still one of the most used payment platforms for eWhoring. Indeed, how to optimise use of various payment platforms is a recurrent topic of conversation in HackForums. For example, various threads discuss mechanisms to bypass PayPal detection. Additionally, in other threads offenders suggest the use Amazon Gift Cards as there are allegedly fewer controls for detecting and closing accounts used for eWhoring.

\subsection{Detect and block saturated images}

The method discussed above, of establishing a hashlist of eWhoring images, could be used by the service providers to detect and block the transmission of offending images. Therefore, this hashlist could be applied not just to the profile pictures used to set up accounts, but also to the images that are distributed to customers, whether that be individual images or frames from cam shows. This method suffers from the same challenges discussed above, i.e. offenders might modify the images to change the hash and bypass detection. Moreover, we note this method would only be effective on communication platforms that do not implement end-to-end encryption.

\subsection{Anonymous reporting}
Offenders are well aware that victims do not report their experience due to the types of activities they are engaged in. Therefore, providing an anonymous reporting platform, such as to the platform being used to generate traffic or provide payments, would enable customers to report if they realised they have been defrauded. This could be practically deployed using readily available software used for whistleblowing, such as GlobalLeaks\footnote{https://github.com/globaleaks/GlobaLeaks} or SecureDrop\footnote{https://github.com/freedomofpress/securedrop}. Competitors who also attempt eWhoring may also make use of such reporting platforms. Reports that include the email address, payment information, or other associated accounts could potentially be shared with other platforms that are facilitating eWhoring. This would enable accounts to be considered for shutdown, and new images to be added to the eWhoring hashlist.

\subsection{Regulate the exchange of alternative currencies}

Exchanging non-fiat currencies, such as selling gift cards for Bitcoin, is currently unregulated. Indeed, underground forums have dedicated boards for currency exchange, which operate in public domains. This makes it more attractive for monetising crime and money laundering. One way to address this is to implement controls to stop the exchange of alternative currencies that are clearly designed to circumvent money-laundering and terrorist financing controls \cite{anderson2018}. 

Those involved in eWhoring may displace to cryptocurrencies for receiving payments, particularly those providing anonymity for transactions (e.g. Monero or Zcash). However, at present this is likely to be counter-productive for actors, since the majority of victims will face a technological barrier. Thus, this intervention focuses on the exchange of income (e.g. through PayPal and Amazon Gift Cards) to cryptocurrencies.

\subsection{Getting caught in the crossfire}

Before implementing any intervention, it is important to consider the impact on the law abiding majority \cite{hancock}. This could involve additional nuisance, such as account verification. However, the impact on some individuals may be greater than others, particularly those legitimately involved in online sex work. For example, it should be considered what would happen if their images were to be stolen, used for eWhoring, and subsequently added to a hashlist, which could inadvertently result in disruption to their own livelihood. 

\subsection{Future innovations}

As with other forms of online crime \cite{smith2003}, eWhoring is likely to change and displace as a result of interventions. In some cases, innovative approaches may reduce some of the harms this crime type currently involves. For example, it may be that actors use photorealistic simulations in the future \cite{RiekWatson2010}. This would reduce the exploitation of real people whose photographs have been obtained and shared without their consent. 

In addition to removing the exploitation of the models, the eWhoring business model may be further legitimised by removing the fraudulent aspects. For example, if the demand for eWhoring is fuelled by loneliness, `social chat bots' \cite{yan2018} may be one positive way forward. These types of innovations are perhaps legitimate ways in which actors currently engaged in eWhoring could utilise their skills, and provide a social benefit, rather than a net harm. 

An overarching tension relates not to eWhoring specifically, but prostitution and the creation and distribution of pornographic content more generally. There are competing views about the commodification of women's bodies, having the personal agency to do so, concerns around exploitation, and women being viewed as sexual objects \cite{weitzer}. We note that photorealistic simulations will create new challenges in this space, such as portraying more extreme encounters, which could create unrealistic expectations for young men and women.

However, future changes, both technical and policy-related, create further criminal opportunities. For example, in the UK under the \textit{Digital Economy Act} 2017, those who wish to access pornographic material will first be required to verify their age. This policy is likely to provide opportunities for scams and frauds. For example, potential customers may believe they are entering credit card details or providing copies of identification documents for a legitimate reason, mandated by government, when they are actually submitting these to a malicious actor who will use or sell them for fraudulent purposes. 

\section{Conclusions}

The eWhoring business model and the distinctive terminology appear to have been developed by innovative online offenders. The largest online underground forum has a dedicated board for this activity, which is rapidly becoming one of the most active. 

While online platforms may provide new opportunities for crime, underground forums also allow researchers to have a fascinating insight into offenders' communications that are often not possible for other crime types. This allows us to analyse the tutorials used for teaching others how to commit new forms of crime, such as eWhoring. 

By qualitatively analysing over 6,500 posts relating to eWhoring tutorials, we identified how this business model operates. Actors mimic models, whose images they have misappropriated, and enter into commercial arrangements with customers, for the purpose of exchanging the images (photographic and video). Script actions include learning the techniques, obtaining images, creating an alias and preparing a backstory, opening accounts, customising images, sourcing traffic, negotiating, receiving payment, sending images, blocking or continuing to milk the customers, and then exchanging funds. 

There are a number of alternative tracks that actors may wish to complement or replace some of the script actions. These include blackmailing, scamming, or sending malware to the customer. They may also become involved in affiliate marketing, obtaining additional income through sending advertisements to potential or actual customers.

eWhoring seems unusual compared to many other crime types, as offenders appear to be cooperating rather than competing. There are likely to be economic reasons for this. The prescriptive approach to the tutorials (including the recommendations not to scam customers) may be tailored to increase the demand for packs of images. Even when images are provided free of charge, there is an expectation that positive reputation will be bestowed. Positive reputation is a valuable commodity on forums that rely on trust \cite{holtsmirnova}. Revenue can also be generated by the advertisements displayed through URL-shortening services. 

Researchers have previously found that the provision of online high yield investment fraud schemes are likely to be driven by the company that built many of the websites \cite{neisius2014}. Similarly, eWhoring is not only enabled, but is likely to be driven, by those who make money from others' offending behaviour. For those stealing intimate images, eWhoring is one way to monetise their activities.  

We suggest a number of potential interventions, noting some rely on the assumption that eWhoring activities are against the terms of use for third party service providers, and they would be willing to enforce these terms. However, the incentives to enforce terms of service are sometimes not apparent to the service provider, and in other cases they face more incentives to retain accounts rather than take them down. For example, having many active accounts depicting attractive women could seem desirable to some service providers. After the adult site Ashley Madison was breached in 2015, it was alleged the majority of accounts purporting to be women on the site were actually `bot' accounts created and operated by the site \cite{newitz15}.

This research has attempted to overcome the significant difficulties associated with this challenging area of research. However, a number of limitations in the research design are identified. First, we capture a static view of eWhoring, while it is likely it has and will change over time, with new opportunities and in response to crime prevention initiatives. Furthermore, it is possible that additional eWhoring methods are not disclosed through the use of tutorials, particularly if they are lucrative. And finally, as this research is exploratory, it is not attempted to quantify the various aspects that have been identified.

We found that crime script analysis is a powerful way to learn about eWhoring, which is a hidden and complex type of fraud. Through this approach we can raise initiatives that may be successful in preventing this crime. Some of the interventions require corporations to understand how their online platforms are being misused. Therefore, by describing it in such a way we can also inform service providers about what is happening on their platforms in a proactive fashion. 

\section*{Acknowledgments}

We thank the anonymous reviewers for their time and insightful comments. We also thank the Cambridge Cybercrime Centre for access to the CrimeBB dataset and the Internet Watch Foundation for their assistance. Finally, we thank our colleagues, particularly Richard Clayton, Daniel Thomas, Alastair Beresford, and Alexander Vetterl, for their invaluable feedback.

This work was supported by the UK Engineering and Physical Sciences Research Council (EPSRC) [grant EP/M020320/1] and by the Comunidad de Madrid (Spain) under the project CYNAMON (P2018/TCS-4566), co-financed by European Structural Funds (ESF and FEDER).

\bibliographystyle{IEEEtran}
\bibliography{IEEEabrv,bibliography}

\end{document}